\documentclass[12pt]{article}
\newdimen\tableauside\tableauside=1.0ex
\newdimen\tableaurule\tableaurule=0.4pt
\newdimen\tableaustep
\def\phantomhrule#1{\hbox{\vbox to0pt{\hrule height\tableaurule width#1\vss}}}
\def\phantomvrule#1{\vbox{\hbox to0pt{\vrule width\tableaurule height#1\hss}}}
\def\sqr{\vbox{%
  \phantomhrule\tableaustep
  \hbox{\phantomvrule\tableaustep\kern\tableaustep\phantomvrule\tableaustep}%
  \hbox{\vbox{\phantomhrule\tableauside}\kern-\tableaurule}}}
\def\squares#1{\hbox{\count0=#1\noindent\loop\sqr
  \advance\count0 by-1 \ifnum\count0>0\repeat}}
\def\tableau#1{\vcenter{\offinterlineskip
  \tableaustep=\tableauside\advance\tableaustep by-\tableaurule
  \kern\normallineskip\hbox
    {\kern\normallineskip\vbox
      {\gettableau#1 0 }%
     \kern\normallineskip\kern\tableaurule}%
  \kern\normallineskip\kern\tableaurule}}
\def\gettableau#1 {\ifnum#1=0\let\next=\null\else
  \squares{#1}\let\next=\gettableau\fi\next}

\def\lsim{\raise0.3ex\hbox{$\;<$\kern-0.75em\raise-1.1ex\hbox{$\sim\;$}}}
\def\gsim{\raise0.3ex\hbox{$\;>$\kern-0.75em\raise-1.1ex\hbox{$\sim\;$}}}

\begin{document}
	
\begin{flushright}
%	NYU-TH-nn/nn/nn \\
	BONN-TH-2005-03\\
	hep-ph/0508150
\end{flushright}
	
\begin{center}
{\bf \large Flavor-singlet hybrid baryons may already have been discovered}\\
\vspace{20mm}

Olaf ~Kittel\\

{\it Physikalisches Institut der Universit\"at Bonn, 
		Nussalle 12,  D-53115 Bonn, Germany \\
\vspace{4mm}
and\\}
\vspace{4mm}
Glennys ~R. ~Farrar

{\it Center for Cosmology and Particle Physics, 
	New York University, NY, NY 10003, USA}

\end{center}

\vspace{25mm}

\begin{abstract}

The splittings between the spin 1/2 and spin 3/2 iso-singlet
baryons $\Lambda_s(1405)$ and $\Lambda_s(1520)$, and their
charmed counterparts $\Lambda_c(2593)$ and $\Lambda_c(2625)$,
have been a theoretical conundrum.  Here we investigate the
possibility that the QCD binding of color octets comprised of
three quarks in a flavor singlet configuration is stronger
than previously envisaged, allowing these states to be
interpreted as hybrids consisting of three quarks plus a valence
gluon $(udsg)$ and $(udcg)$.  A fit of their mass separation
allows the mass prediction of the strange and charmed flavor octet 
and decuplet hybrid baryons and the prediction of the mass separation
of the beauty hybrids.
Such hybrid states come in parity-doubled pairs with the
even parity state lighter by about $300$~MeV.  Existing data
accommodates either parity assignment for the observed states and
the existence of the required unobserved partners
at either higher or lower mass. 
We discuss difficulties with and
strategies for observing the other states under the two cases. A
corollary of the strong-binding-in-flavor-singlet-channel hypothesis 
is that the H-dibaryon may be
very long lived or stable with $m_H \lsim 2$~GeV.

%{\bf b-quarks, mesons, hybridmesons }

\end{abstract}

\newpage

\section{Introduction}

The spectrum of baryons between $1$~GeV and $2$~GeV has been accurately
measured experimentally \cite{1.0} and in most cases detailed partial
wave analyses have lead to the determination of their spins and
parities.  Nearly all known baryon resonances are well-described as
three-quark states, however the isosinglets $\Lambda_s(1405)$ 
(total angular momentum $J=1/2$) and $\Lambda_s(1520)$ ($J=3/2$) 
are problematic in this interpretation \cite{Dal}.  
If they are orbital momentum $L=1$ states in the three-quark model, their large energy splitting of $115$~MeV cannot be explained 
by  QCD calculations \cite{CapIsg}, which predict the two states
to be nearly degenerate. Calculations which include spin-orbit
interactions even predict an inversion of the masses \cite{CloDal}, 
contrary to fact.
An alternative suggestion is that the $\Lambda_s(1405)$ 
is a $\bar K$-$N$ $s$-wave bound state \cite{ArMa}.  
Since the $\bar K$-$N$ $p$-wave system would not bind,
this interpretation does not provide any $J=3/2$ partner, so the $\Lambda_s(1520)$ is taken to be a $L=1$ three-quark state, 
and its $J=1/2$ partner is assumed to be hiding, close in mass 
according to \cite{CapIsg}. However this is also problematic, since in 
$\bar K$-$N$ experiments this mass region is well explored without 
evidence of such a resonance \cite{Dal}.

QCD predicts other composite particles, color singlet states not only
involving quarks but also ``constituent'' gluons \cite{BaClVi}. The
phenomenology of these baryons, called hybrids, was developed by
Barnes and Close \cite{BaClo, BarClos}, Golowich et al. \cite{GoHa}
and others. These works use the bag model and the potential model to
suggest that the lightest hybrid masses are below $2$~GeV.  Although
there is evidence in favor of some candidates \cite{Mey}, no hybrids
have yet been confirmed.  One problem is that the mixing of hybrids
with conventional states makes them hard to discriminate from normal
states.  Since the quantum numbers of hybrid states sometimes coincide
with those of known particles, it is also possible that some octet and
decuplet states identified as conventional three-quark baryon states could
be hybrids \cite{CaHa, Karl, FEClose}.

Here we explore the possibility that the observed $J= 1/2$
and $J = 3/2$ isosinglet baryons 
$\Lambda(1405)$ and $\Lambda(1520)$ are actually hybrid
baryon states \cite{Far}.
To test the hypothesis, we calculate their
mass splittings in the hybrid picture and 
	find the correct amount ($\approx 100$~MeV) and
	ordering ($m_{3/2}>m_{1/2}$).
%Under the hybrid assumption, we can also make mass predictions for the 
%octet and decuplet hybrid baryons.
Since the direct determination of the parity of $\Lambda(1405)$ is 
experimentally delicate and yet unclear \cite{Dal}, we consider 
both possibilities for its parity. 
%see section \ref{The parity of Lambda(1405)}.
A constituent gluon combining with three quarks in
an $L=0$ orbital ground state can form both parity even and parity odd
states, with the latter about $300$~MeV lighter \cite{Closebook}.  
Remarkably, both parity assumptions for the hybrids 
are compatible with existing data.  
%We discuss options for removing this ambiguity.

The paper is organized as follows. 
In Section \ref{Definitions and formalism} we review the general
structure of hybrid baryon wave functions. We define the degeneracy lifting
hyperfine interaction using effective  hyperfine couplings between the
constituents.
In Section \ref{Mass splittings of hybrid baryons}
we determine the  quark-quark couplings from ordinary baryon 
mass splittings 
%in Section 
%\ref{Quark-quark couplings for baryons}.
%The interaction strength is determined by
%an effective 1 gluon exchange between quarks and between a quark
%and a gluon.  
and calculate the flavor singlet hybrid mass splittings.
%, discover they have
%the correct ordering $m_{3/2}>m_{1/2}$,  
We then use the observed flavor singlet
splittings to fix the quark-gluon effective hyperfine coupling.
This allows predictions to be made for octet and decuplet hybrid
baryon mass splittings. 
%which can serve as a test of this model
%when the relevant regions of mass are explored.  
In Section \ref{Mass splittings in the mesonic sector}
we obtain the hyperfine couplings in the mesonic
sector and relate them to the values found for baryonic states. 
%in Section \ref{Mass splittings of hybrid baryons}.
In Sections
\ref{The parity of Lambda(1405)} and \ref{A low lying dihyperon
?} two implications of this model are discussed -- parity
doubling and a possible light H-dibaryon.  Section
\ref{Summary and Conclusion} gives a summary of our results and
conclusions.

A preliminary description of this work was reported in 
\cite{kf:Lam1405}.  Since then, the experimental viability of a
stable H has been established \cite{f:stableH, fz:nucStab,
fz:binding}, making this scenario more credible.
In addition, the possible discovery of pentaquark states 
in the $1.5$~GeV range underlines the need to be open-minded regarding
the validity of naive QM dynamical assumptions.

\section{\label{Definitions and formalism}
                Definitions and formalism}

\subsection{\label{Wave functions of hybrid baryons}
                   Wave functions of hybrid baryons}

A systematic group theoretical classification of the hybrid wave 
functions is possible, as the quarks have to
obey the Pauli principle.
We consider systems of three quarks in the orbital
ground state which have up to three
different quark flavors $u$, $d$ and $i$ with $i=s,c,b$. The flavor group 
is $SU(3)_F$ if flavor symmetry is assumed, i.e. the mass differences 
between quarks of different flavors are neglected. 
The groups for color and spin are $SU(3)_C$ and $SU(2)_S$,
respectively. A single quark is then
a 18-dimensional representation of the direct product
group $SU(3)_C \times SU(3)_F \times SU(2)_S$.
We reduce the direct product of the three quarks in
irreducible representations of $SU(18)$ with the help of Young
tableaux \cite{Sche}
\begin{eqnarray} \label{SU(18) decomposition}
  {\tableau{1} \atop{\bf 18}}
  {\times  \atop }
  {\tableau{1} \atop{\bf18}}
  {\times \atop }
  {\tableau{1} \atop{\bf18}} {= \atop }
    {\tableau{3} \atop{\bf 1140}} {+ \atop }
    {\tableau{2 1} \atop{\bf 1938}} {+ \atop }
    {\tableau{2 1} \atop{\bf 1938}} {+ \atop }
    {\tableau{1 1 1} \atop{\bf 816}} \, .
\end{eqnarray}
%We label the resulting irreducible representations by their dimensions. 
%The completely antisymmetric representation is the {\bf 816}.
All physical $qqq$-states which obey the Pauli principle are members of 
the  completely antisymmetric representation {\bf 816}.
The decompositions of the multiplet into representations
of $SU(3)_C$, $SU(3)_F$ and $SU(2)_S$ are given in Tab.~\ref{Decomposition}.
\begin{table}
 \caption{\label{Decomposition}Decomposition of $SU(18)$ in
                  representations of color, flavor and spin.}
\[
  \begin{tabular}{|l||c|c|c||c|}  \hline
$SU(18)$&$SU(3)_C$ & $SU(3)_F$ & $SU(2)_S$ & \# of these \\ \hline
                                                            \hline
        & $  { \tableau{1 1 1} \atop\textstyle{\bf 1}}$
        & $  { \tableau{2 1}   \atop\textstyle{\bf  8}}$
        & $  { \tableau{2 1}   \atop\textstyle{\bf 2}}$  &16 \\ \cline{2-5}
        & $  { \tableau{1 1 1} \atop\textstyle{\bf 1}}$
        & $  { \tableau{3}     \atop\textstyle{\bf 10}}$
        & $  { \tableau{3}     \atop\textstyle{\bf 4}}$  &40 \\ \cline{2-5} \cline{2-5}
        & $  { \tableau{3}     \atop\textstyle{\bf 10}}$
        & $  { \tableau{2 1}   \atop\textstyle{\bf  8}}$
        & $  { \tableau{2 1}   \atop\textstyle{\bf  2}}$  &160\\ \cline{2-5}
 $ {\tableau{1 1 1} \atop\textstyle{\bf 816}}$
        & $  { \tableau{3}     \atop\textstyle{\bf 10}}$
        & $  { \tableau{1 1 1} \atop\textstyle{\bf  1}}$
        & $  { \tableau{3}     \atop\textstyle{\bf  4}}$  & 40\\ \cline{2-5} \cline{2-5}
        & $  { \tableau{2 1}   \atop\textstyle{\bf  8}}$
        & $  { \tableau{2 1}   \atop\textstyle{\bf  8}}$
        & $  { \tableau{2 1}   \atop\textstyle{\bf  2}}$  &128\\ \cline{2-5}
        & $  { \tableau{2 1}   \atop\textstyle{\bf 8}}$
        & $  { \tableau{2 1}   \atop\textstyle{\bf 8}}$
        & $  { \tableau{3}     \atop\textstyle{\bf 4}}$  &256\\ \cline{2-5}
        & $  { \tableau{2 1}   \atop\textstyle{\bf 8}}$
        & $  { \tableau{3}     \atop\textstyle{\bf 10}}$
        & $  { \tableau{2 1}   \atop\textstyle{\bf  2}}$  &160\\ \cline{2-5}
        & $  { \tableau{2 1}   \atop\textstyle{\bf  8}}$
        & $  { \tableau{1 1 1} \atop\textstyle{\bf  1}}$
        & $  { \tableau{2 1}   \atop\textstyle{\bf  2}}$  & 16\\ \hline
   \end{tabular}
  \]
\end{table}
The gluon is in the color octet representation of $SU(3)_C$. The hybrid
has to be a color singlet, thus the quarks have to be in the
complex conjugate representation, which is again an octet.
We will distinguish the $qqq$ color octet states in the lower half of
Tab.~\ref{Decomposition} by the shorthand
$^{spin}flavor$: $ \bf{ ^2 8}, \:^4 8,\: ^2 10,\: ^2 1$.
There are altogether 70 color octet states with mixed symmetry
which form a $SU(6)$ representation of $SU(3)_F \times SU(2)_S$
\begin{eqnarray}  \label{SU(6) reduction}
  {\tableau{1} \atop {\bf 6}}
  {\times  \atop }
  {\tableau{1} \atop{\bf6}}
  {\times \atop }
  {\tableau{1} \atop{\bf6}} {= \atop }
    {\tableau{3} \atop{\bf 56}} {+ \atop }
    {\tableau{2 1} \atop{\bf 70}} {+ \atop }
    {\tableau{2 1} \atop{\bf 70}} {+ \atop }
    {\tableau{1 1 1} \atop{\bf 20}} \, .
\end{eqnarray}
A Young tableau of the form $\tableau{2 1}$ is mixed symmetric 
because the wave function which is assigned to the tableaux 
is symmetric (or antisymmetric) only under the exchange of two quarks.
We choose them to be the first and second quark and will use this
convention throughout. Since this choice is arbitrary, we give 
the wave functions in flavor space in Appendix
\ref{Mixed symmetry functions}.
We label the wave function $\varphi_{MS}$ for mixed symmetric states 
and $\varphi_{MA}$ for mixed antisymmetric states.
A totally symmetric (totally antisymmetric)
function $\phi_{S(A)}$, under the interchange of any two quarks,
can be built out of mixed symmetric functions
$\varphi_{MS(A)}$ and $\varphi_{MS(A)}'$ \cite{Closebook}
 \begin{eqnarray}
      \phi_S &=& \displaystyle
                \frac{1}{\sqrt{2}}
      (\varphi_{MS}\varphi_{MS}' + \varphi_{MA}\varphi_{MA}'), \\
      \phi_A &=& \displaystyle
                \frac{1}{\sqrt{2}}
      (\varphi_{MS}\varphi_{MA}' - \varphi_{MA} \varphi_{MS}' ).
  \end{eqnarray}
Mixed symmetric functions $\phi_{MS}$ and 
mixed antisymmetric functions $\phi_{MA}$ are \cite{Closebook}
  \begin{eqnarray}
      \phi_{MS} &=& \frac{1}{\sqrt{2}}
      (-\varphi_{MS}\varphi_{MS}' + \varphi_{MA}\varphi_{MA}'), \\
      \phi_{MA} &=& \frac{1}{\sqrt{2}}
      (\varphi_{MS}\varphi_{MA}' + \varphi_{MA} \varphi_{MS}' ).
 \end{eqnarray}
The four color octet wave functions, which lie in the {\bf 816} 
of $SU(18)$, are then \cite{Closebook}
\begin{eqnarray} \label{48 wave function}
 {\bf^4 8}:~\psi_A &=& \displaystyle
        \frac{1}{\sqrt{2}}
                    \big(
                        c_{MS}f_{MA}- c_{MA}f_{MS}
                    \big)  s_{S}, \\ \label{210 wave function}
 {\bf^2 10}:~\psi_A &=&   \displaystyle
         \frac{1}{\sqrt{2}}(c_{MS}s_{MA}-c_{MA}s_{MS})f_{S},\\
 {\bf^2 8}:~\psi_A &=& \displaystyle
         \frac{1}{2}
                \Big[
                       c_{MS}(f_{MA}s_{MS}+f_{MS}s_{MA})
                      -c_{MA} (f_{MA}s_{MA}-f_{MS}s_{MS} )
                \Big],\\ \label{The singlet wave function}
 {\bf^2 1}:~\psi_A &=& \displaystyle
 \frac{1}{\sqrt{2}}(c_{MA}s_{MA} + c_{MS}s_{MS})f_{A}.
 \label{48 wave function d}
  \end{eqnarray}
with $\varphi = f,s,c$ for  flavor, spin and the color 
three quark wave function, respectively.
We will include flavor breaking and work in the approximation
that isospin is exact. In that case, we distinguish between 
isospin singlet $(I=0)$ and isospin triplet $(I=1)$ hybrid baryons
\begin{eqnarray}
	{\bf ^4 8} (I=0,1), \quad 
	{\bf ^2 10}(I=1),\quad 
	{\bf ^2 8} (I=0,1),\quad 
	{\bf ^2 1} (I=0).
\end{eqnarray}
These states combine with the gluon,
which thus gains effective mass \cite{BarClos, GoHa} 
and a third degree of freedom, the spin zero state.
The confined gluon is classified by a TE or TM mode \cite{Closebook, TaDo}, 
with total angular momentum $J$ and parity $P$
 \begin{eqnarray}
	  TE: \; J^P=1^{+}, 2^{-}, \dots \quad \quad
	  TM: \; J^P=1^{-}, 2^{+}, \dots \; .
   \end{eqnarray}
In the bag model, the TE $J^P=1^{+}$ mode is
estimated to be about $300$~MeV lighter 
than the TM $J^P=1^{-}$ mode \cite{Closebook}.

\subsection{\label{Interaction Hamiltonian}
                   Interaction Hamiltonian}

We divide the hyperfine interaction Hamiltonian $V_{hyp}$ 
into interactions only between the quarks
$V_{qq}$, and interactions between  the quarks
and the  constituent gluon $V_{qg}$
\begin{eqnarray} \label{Vhyp}
 V_{hyp} & = & V_{qq} + V_{qg}.
\end{eqnarray}
With the effective hyperfine couplings $ \kappa_{ij}$
between the quarks, we define the interaction
\begin{eqnarray}\label{Vqq}
	V_{qq} &=& -\sum_{i<j}\kappa_{ij} 
		\,{\bf S}^i \cdot {\bf S}^j \; {\bf F}^i \cdot {\bf F}^j,
\end{eqnarray}
which has the color-spin structure of 1-gluon exchange \cite{GoHa}.
The spin matrices ${\bf S}^i$ and color matrices ${\bf F}^i$ 
for the $i$th quark are
\begin{eqnarray}
          S^i_m = \frac{1}{2}\sigma_m,  \quad
          F^i_a = \frac{1}{2}\lambda_a,
\end{eqnarray}
where $\sigma_m$, $m = 1,2,3$, and 
$\lambda_a$,  $a = 1,\dots,8$, are the Pauli and the Gell-Mann matrices 
\cite{Closebook}, and the matrix products are defined by
\begin{eqnarray}
	{\bf S}^i \cdot {\bf S}^j =\sum_{m=1}^3 S^i_m \, S^j_m, \quad  
	{\bf F}^i \cdot {\bf F}^j =\sum_{a=1}^8 F^i_a \, F^j_a.
\end{eqnarray}
The effective one-gluon ansatz leads to the quark-gluon interaction 
\begin{eqnarray}\label{Vqg}
	V_{qg} &=& - \sum_{i} \kappa_{ig} 
		\,{\bf S}^i \cdot {\bf S}^g \;{\bf F}^i \cdot {\bf F}^g,
\end{eqnarray}
where ${\bf S}^g$ and ${\bf F}^g$ are the gluon spin and color matrices. 
The  Hamiltonian is then
\begin{eqnarray} \label{the interaction Hamiltonian}
	V_{hyp} &=& -\sum_{i<j}\kappa_{ij} 
		\,{\bf S}^i \cdot {\bf S}^j \; {\bf F}^i \cdot {\bf F}^j 
	- \sum_{i} \kappa_{ig} 
		\,{\bf S}^i \cdot {\bf S}^g \;{\bf F}^i \cdot {\bf F}^g.
\end{eqnarray}
We label the effective coupling between two light quarks by $\kappa$ 
and those between a light and a heavy quark by $\kappa_i$, 
with $i=s, c, b$ the index for the heavy quark. 
The coupling between a gluon and a light quark is labeled by $\kappa_g$ 
and those between a gluon and a heavy quark is labeled by $\kappa_{ig}$.
% thus breaking  $SU(3)_F$ symmetry. 
The hierarchy of the $\kappa$'s follows from the form of the 
``Fermi-Breit-interaction'' in QCD for single gluon exchange \cite{FEClose}
  \begin{eqnarray}
	  V^{ij} \propto \frac{{\bf F}^i\cdot {\bf F}^j \; 
							{\bf S}^i\cdot {\bf S}^j} {m_i \;m_j}.
	\label{hypcoeff}
\end{eqnarray}
The effective coupling $\kappa$ would therefore be 
inversely proportional to the product of the masses of the interacting
particles
\begin{eqnarray} \label{kappa mass relation}
  \frac{ \kappa_i}{\kappa_j}= \frac{m_j}{m_i}.
\end{eqnarray}
Moreover, the hyperfine coefficient (\ref{hypcoeff}) is the product of
the color magnetic moments. Thus
replacing a light quark color magnetic moment with a gluon
magnetic moment we could determine $\kappa_g$ from $\kappa$. Since this is 
the same replacement independent of quark flavor, we expect 
 \begin{eqnarray} \label{kappa relation}
       \frac{\kappa}{\kappa_{g}}=
    \frac{\kappa_s}{\kappa_{sg}}=
    \frac{\kappa_c}{\kappa_{cg}}.
 \end{eqnarray}
These relations reduce the number of parameters in the fit of hybrid
baryons and makes it more predictive.
We determine the coupling strengths 
$\kappa$ and $\kappa_i$ from ordinary baryon mass splittings in Section 
\ref{Quark-quark couplings for baryons} and $\kappa_g$ and $\kappa_{ig}$
in Section \ref{Mass splittings of the flavor singlet hybrid baryons}.

\subsection{\label{Quark-Quark Interactions}
                   Quark-quark interactions}

We calculate the matrix elements $<V_{qq}>$~(\ref{Vqq})
of the interaction Hamiltonian~(\ref{Vhyp}),
including flavor breaking and assuming that isospin is
conserved. An operator $O$ for a baryon with two light quarks 
$q=u,d$ and one heavy quark $i=s,c,b$ has then the structure 
$O = O^S +\epsilon~O^{12} $. The part $O^S$ is flavor symmetric 
and symmetric under the interchange of any two quarks. The part $O^{12}$ 
is isospin symmetric and symmetric under the interchange of the light
quarks 1 and 2 only. 
%in the basis (see Appendix \ref{Definitions of the mixed symmetry
%functions}) in which the light quarks are chosen to be 1 and 2.
%In the limit of unbroken flavor symmetry, $\epsilon$ goes to zero. 
The measure for flavor breaking is the mass difference between the heavy
and the light quark. 
In the limit of unbroken flavor symmetry, $\epsilon$ goes to zero.
%$\epsilon$ is proportional to $ m_i - m_q$. 
Such a  decomposition of $V_{qq}$ leads to
\begin{eqnarray}\label{qq interaction}
	V_{qq} &=& - \kappa_3 \, \sum_{i<j} {\bf S}^i\cdot {\bf S}^j 
			\,{\bf F}^i\cdot {\bf F}^j - 
(\kappa - \kappa_3) \,{\bf S}^1\cdot {\bf S}^2 \,{\bf F}^1\cdot{\bf F}^2.
\end{eqnarray}
The first term of~(\ref{qq interaction}) is completely symmetric. 
Its evaluation  for exact flavor symmetry leads to \cite{BuFaPu}
 \begin{eqnarray} \label{Expectation value of OA}
 < \sum_{i<j} {\bf S}^i\cdot {\bf S}^j \,{\bf F}^i\cdot {\bf F}^j>
         &=&  \frac{21}{16}  - \frac{1}{8}C_C^{qqq}
          - \frac{1}{4}C_F^{qqq} - \frac{1}{12}C_S^{qqq},
 \end{eqnarray}
with $C_{C,F,S}^{qqq}$ the Casimir operators of the three quarks for
color, flavor and spin, respectively, as defined in Appendix 
\ref{Casimir operators}. Alternatively, this term can  be evaluated using 
the $SU(6)$ Casimir operators $C_6^{qqq}$
of color and spin of the three quarks \cite{Jaffe}
\begin{eqnarray}
	< \sum_{i<j} {\bf S}^i\cdot {\bf S}^j \,{\bf F}^i\cdot {\bf F}^j>
         &=&  \frac{1}{32}C_6^{qqq} - \frac{1}{12}C_S^{qqq}
         - \frac{1}{8}C_C^{qqq} - \frac{3}{2}.
  \end{eqnarray}
The definition and values of $C_6$ in various representations 
may be found in \cite{Jaffe2}.
We give the values of 
$ <\sum_{i<j} {\bf S}^i\cdot {\bf S}^j \, {\bf F}^i\cdot {\bf F}^j> $
for the color octet three quark states in the first part of 
Tab.~\ref{The quark-quark interactions for color octets}.
\begin{table}[h]
  \caption{\label{The quark-quark interactions for color octets}
	  Expectation values of $ O^A=\sum_{i<j} {\bf S}^i\cdot {\bf S}^j  \,
	  {\bf F}^i\cdot {\bf F}^j$
    and  $ O^{12}={\bf S}^1\cdot {\bf S}^2 \,{\bf F}^1\cdot {\bf F}^2$ }
\[
  \begin{array}{|c||c|c|c|c|} \hline
SU(3)_C & {\bf8} & {\bf8} & {\bf8} &  {\bf8} \\ \hline
SU(3)_F & {\bf8} & {\bf8} & {\bf10}&  {\bf1} \\ \hline
SU(2)_S & {\bf2} & {\bf4 }& {\bf2} &  {\bf2} \\ \hline \hline
<O^A>   &  1/8   & -1/8   &  -5/8  &     7/8 \\ \hline
  \end{array} \;\;\;\;\;\;\;\;\;
  \begin{array}{|c||c|c|c|c|} \hline
SU(3)_C & {\bf\bar{3}} & {\bf\bar{3}} & {\bf6} & {\bf6} \\ \hline
SU(2)_I &    {\bf1}    &    {\bf3}    & {\bf1} & {\bf3} \\ \hline
SU(2)_S &    {\bf1}    &    {\bf3}    & {\bf3} & {\bf1} \\ \hline \hline
<O^{12}>&   1/2        & -1/6         &   1/12 &  -1/4  \\ \hline
  \end{array}
  \]
\end{table}

The second term of~(\ref{qq interaction}) is symmetric 
under interchange of quark 1 and 2.
Using isospin rather than flavor, we can 
modify~(\ref{Expectation value of OA}) for this case and find
\begin{eqnarray} \label{Expectation value of O12}
 <{\bf S}^1\cdot {\bf S}^2\,{\bf F}^1\cdot {\bf F}^2> &=&
     \frac{2}{3}-\frac{1}{8} C_C^{12} -
                 \frac{1}{4} C_I^{12} -
                 \frac{1}{12}C_S^{12}.
\end{eqnarray}
$C_{C,I,S}^{12}$ are the Casimir operators of the quark 1 and 2 
for color, isospin and spin, respectively. We give the values for
$ <{\bf S}^1\cdot {\bf S}^2\, {\bf F}^1\cdot {\bf F}^2> $ 
for all possible antisymmetric
representations in flavor, color and isospin of the diquark in the second
part of Tab.~\ref{The quark-quark interactions for color octets}.

\subsection{\label{Quark-gluon interactions}
                   Quark-gluon interactions}

The quark-gluon interaction~(\ref{Vqg}) also
decomposes into two terms
 \begin{eqnarray} \label{quark-gluon decomposition}
     V_{qg} &=&  -\kappa_{3g}\sum_i {\bf S}^i\cdot {\bf S}^g\,
		{\bf F}^i\cdot {\bf F}^g  \nonumber \\
					&&  - (\kappa_{g}-\kappa_{3g})
					({\bf S}^1\cdot {\bf S}^g\,{\bf F}^1\cdot {\bf F}^g+
					 {\bf S}^2\cdot {\bf S}^g\,{\bf F}^2\cdot {\bf F}^g).
  \end{eqnarray}
The first term is completely symmetric under interchange of any
pair of quarks and the second term is symmetric only under interchange of
quark 1 and 2. We list our results for the flavor singlet and the results for
the light octets and decuplets \cite{BarClos} in Tab. 
\ref{Values for sumiSiSgFiFg}.
\begin{table}[h]
 \caption{\label{Values for sumiSiSgFiFg}
    Values for $<\sum_{i} {\bf S}^i\cdot {\bf S}^g\,{\bf F}^i\cdot {\bf F}^g>$
   and $<{\bf S}^1\cdot {\bf S}^g\,{\bf F}^1\cdot {\bf F}^g>$}
\[
   \begin{tabular}{|c||c|c|c||c|c||c|c||c|c||} \hline
 $^{spin} flavor$ & \multicolumn{3}{c||}{$\bf^48$} &
                    \multicolumn{2}{c||}{$\bf^210$}&
                    \multicolumn{2}{c||}{$\bf^28$}&
                    \multicolumn{2}{c||}{$\bf^21$} \\ \hline
  total J & 5/2&3/2&1/2&3/2&1/2&3/2&1/2&3/2&1/2  \\ \hline \hline
$<\sum_{i} {\bf S}^i\cdot {\bf S}^g\,{\bf F}^i\cdot {\bf F}^g>$&
    -3/2& 1&5/2&0&0&-1/2&1&-1 &2\\ \hline \hline
        &   \multicolumn{3}{c||}{I=1}&
            \multicolumn{2}{c||}{I=1}&
            \multicolumn{2}{c||}{I=1}&
            \multicolumn{2}{c||}{I=0} \\ \cline{2-10}
$<{\bf S}^1\cdot {\bf S}^g\,{\bf F}^1\cdot {\bf F}^g>$
  & -3/8&1/4&5/8&-1/4&1/2&-1/4&1/2&0&0 \\  \cline{2-10}
    &       \multicolumn{3}{c||}{I=0}&
            \multicolumn{2}{c||}{}&
            \multicolumn{2}{c||}{I=0}&
            \multicolumn{2}{c||}{} \\ \cline{2-4}  \cline{7-8}
    &-5/8& 5/12&25/24&\multicolumn{2}{c||}{} &
                 0&0& \multicolumn{2}{c||}{}\\ \hline
   \end{tabular}
  \]
\end{table}

For the flavor octets and decuplets which contain only light quarks
$u$  and $d$, the second term of~(\ref{quark-gluon decomposition}) 
vanishes because $\kappa_{g}=\kappa_{3g}$.
For hybrids containing one heavy quark $i=s,c,b$ we use the symmetry 
of the wave function under interchange of quark 1 and 2
\begin{eqnarray} \label{symmetry 12}
	<{\bf S}^1\cdot {\bf S}^g\,{\bf F}^1\cdot {\bf F}^g> &=& 
	<{\bf S}^2\cdot {\bf S}^g\,{\bf F}^2\cdot {\bf F}^g>,
\end{eqnarray}
and write
\begin{eqnarray}
  \lefteqn{
           <{\bf S}^1\cdot {\bf S}^g\,{\bf F}^1\cdot {\bf F}^g>
          }      \nonumber \\
               &=&  \frac{1}{4}
                       \left[
                            ({\bf S}^{1}+{\bf S}^{g})^2
                           -({\bf S}^{1})^2-({\bf S}^{g})^2
                        \right]
                       \left[
                            ({\bf F}^{1}+{\bf F}^{g})^2
                           -({\bf F}^{1})^2-({\bf F}^{g})^2
                        \right]   \nonumber   \\
               &=&  \frac{1}{4}
                       \left[
                              ({\bf S}^{1}+{\bf S}^{g})^2 -\frac{11}{4}
                      \right]
                       \left[
                              ({\bf F}^{1}+{\bf F}^{g})^2 -\frac{13}{3}
                       \right].  \label{xy}
\end{eqnarray}
What remains is to determine the spin and color representations of the
gluon--first quark state (${\bf S}^1+{\bf S}^g$ and ${\bf F}^1+{\bf F}^g$) 
of each hybrid baryon. In order to expand the color and spin wave function of
each hybrid (with the help of $SU(2)$ \cite{1.0} and $SU(3)$ \cite{Clebsch}
Clebsch Gordan coefficients)
into  gluon--first quark and second quark--third quark color and spin
wave functions, we need to know the spin and color representations of
the diquark. The 1-2 antisymmetric part of each hybrid wave function has to
fulfill two constraints. Firstly, the flavor part must be
$f_{MS}$ (or $f_S$) for isotriplets and $f_{MA}$ (or $f_A$) for isosinglets.
Secondly, the wave function must be antisymmetric under interchange of quark
1 and 2. The wave functions for the $\bf ^48$ hybrids result immediately
from (\ref{48 wave function})
\begin{eqnarray}
  {\bf ^48}(I=1)&:& f_{MS}c_{MA}s_S, \\
  {\bf ^48}(I=0)&:& f_{MA}c_{MS}s_S.
\end{eqnarray}
For the other hybrids states  
(\ref{210 wave function})-(\ref{The singlet wave function}), 
the wave function of each hybrid
may be a linear combination (l.c.) of the following functions
\begin{eqnarray}
{\bf ^21}(I=0)&:& {\rm l.c. \;\; of}\;\; f_Ac_{MS}s_{MS} \;\;{\rm and} \;\; f_Ac_{MA}s_{MA}, \\
 {\bf ^210}(I=1)&:& {\rm l.c. \;\; of}\;\; f_Sc_{MS}s_{MA} \;\;{\rm and} \;\; f_Sc_{MA}s_{MS}, \\
  {\bf ^28}(I=1)&:& {\rm l.c. \;\; of}\;\; f_{MS}c_{MS}s_{MA}\;\;{\rm and} \;\; f_{MS}c_{MA}s_{MS}, \\
  {\bf ^28}(I=0)&:& {\rm l.c. \;\; of}\;\; f_{MA}c_{MS}s_{MS}\;\;{\rm and} \;\; f_{MA}c_{MA}s_{MA}.
\end{eqnarray}
All the above parts of the wave functions are eigenfunctions
of the operator
\begin{eqnarray}
O^{12}&=& -(\kappa-\kappa_3)\,{\bf S}^1\cdot{\bf S}^2\,{\bf F}^1\cdot{\bf F}^2
- 2(\kappa_g-\kappa_{3g})\,{\bf S}^1 \cdot{\bf S}^g\,{\bf F}^1\cdot {\bf F}^g,
\end{eqnarray}
which is the 1-2 symmetric part of our interaction Hamiltonian
$V_{hyp}$ (\ref{Vhyp}). We assume that the diquark is in a state
in which the energy is minimal, i.e., in which $O^{12}$ is minimized.
The wave functions with minimal energy are
\begin{eqnarray}
  {\bf ^21}(I=0)&:&  f_Ac_{MA}s_{MA}, \\
 {\bf ^210}(I=1)&:&  f_Sc_{MA}s_{MS}, \\
  {\bf ^28}(I=1)&:&  f_{MS}c_{MA}s_{MS}, \\
  {\bf ^28}(I=0)&:&  f_{MA}c_{MA}s_{MA}.
\end{eqnarray}
We list the resulting values for 
$<{\bf S}^1\cdot {\bf S}^g\,{\bf F}^1\cdot {\bf F}^g>$
in Tab.~\ref{Values for sumiSiSgFiFg},
and values for $<{\bf S}^1\cdot {\bf S}^2\,{\bf F}^1\cdot {\bf F}^2>$ 
can be taken from the second part of 
Tab.~\ref{The quark-quark interactions for color octets}.

\section{\label{Mass splittings of hybrid baryons}
                Mass splittings of hybrid baryons}

\subsection{\label{Quark-quark couplings for baryons}
                   Quark-quark couplings for baryons}

In order to find values for the effective  couplings
$\kappa$, $\kappa_s$ and $\kappa_c$ we fit 
the mass splittings of the baryons 
$\Sigma_i^*$, $\Sigma_i$ and $\Lambda_i$, labeled by the flavor 
index of the heavy quark $i=s,c,b$.
In the case of exact $SU(3)_F \times SU(2)_S$ symmetry,
the isospin multiplets $\Sigma_i^*$, $\Sigma_i$ and $\Lambda_i$ 
(for $i$ fixed) would be members of the totally symmetric 56-dimensional 
representation, see~(\ref{SU(6) reduction}).
By switching on the mass difference between the heavy quark $i$ and the
light quarks $q$ we break $SU(3)_F$. The
values of $\kappa$, $\kappa_s$ and $\kappa_c$ can thus be determined
by the experimentally observed mass separations of the isospin multiplets
given in Tab.~\ref{The particles used to determine kappai}.  
Calculating the quark-quark interaction~(\ref{qq interaction}),
we find for these mass splittings \cite{Closebook}
\begin{eqnarray}
  E(\Sigma_i^*) - E(\Sigma_i)&=& \kappa_i,   \\
 E(\Sigma_i) - E(\Lambda_i) &=& \frac{2}{3}\kappa - \frac{2}{3}\kappa_i, \\
  E(\Sigma_i^*) - E(\Lambda_i)&=& \frac{2}{3}\kappa + \frac{1}{3}\kappa_i.
\end{eqnarray}
With increasing mass of the heavy quark $i$, $\kappa_i$ decreases
(\ref{kappa mass relation}) so that the $\Sigma_i^*-\Sigma_i$
splittings decrease, the $\Sigma_i -\Lambda_i$ splittings increase
and the $\Sigma_i^*- \Lambda_i$ splittings decrease.

In addition, we can predict the order of the $\Sigma_b^*$, $\Sigma_b$
and $\Lambda_b$ mass splittings, which have not yet been measured. 
Their mass splittings depend on $\kappa_b$ which can be estimated 
with $m_c=1.25$~GeV, $m_b=4.25$~GeV \cite{1.0} and 
relation (\ref{kappa mass relation})
\begin{eqnarray}
	\kappa_b= \frac{m_c}{m_b} \kappa_c =18~{\rm MeV}.
\end{eqnarray}
We give the mass splittings of the beauty baryons 
in Tab.~\ref{The particles used to determine kappai}.
\begin{table}[h]
 \caption{\label{The particles used to determine kappai}
                The baryons used to fix the $\kappa_i$ and predicted
					 splittings of beauty baryons}
\[
   \begin{tabular}{|r|c|c|c|c|c|c|c|c|} \hline
  $\kappa_i$&  baryon&spin&isospin&flavor&content&
          $\Delta M~[\rm MeV]$&fit&$\kappa_i~[\rm MeV]$\\ \hline\hline
   $\kappa$ &$\Delta(1232)$& 3/2&3/2& {\bf10}&qqq& $\Delta-N=293$&293&293   \\
            &   $N(939)$   & 1/2&1/2& {\bf8} &qqq& &&                \\
                              \hline
  $\kappa_s$&$\Sigma_s^*(1385)$&3/2&1&{\bf10}&qqs&$\Sigma_s^*-\Sigma_s=192$&182&182\\
            &$\Sigma_s(1193)$& 1/2&1& {\bf8} &qqs&$\Sigma_s-\Lambda_s=77 $&74&\\
            &$\Lambda_s(1116)$& 1/2&0& {\bf8}&qqs&$\Sigma_s^*-\Lambda_s=269 $&256&\\
                              \hline
  $\kappa_c$&$\Sigma_c^*(2520)$&3/2&1&{\bf10}&qqc&$\Sigma_c^*-\Sigma_c=65 $&60&60\\
            &$\Sigma_c(2455)$& 1/2&1& {\bf8 }&qqc&$\Sigma_c-\Lambda_c=170 $&155& \\
            &$\Lambda_c(2285)$& 1/2&0& {\bf8}&qqc&$\Sigma_c^*-\Lambda_c=235 $&215& \\
                              \hline\hline
  $\kappa_b$&$\Sigma_b^*(?)$& 3/2&1& {\bf10}&qqb&$\Sigma_b^*-\Sigma_b=? $&18&18\\
            &$\Sigma_b(?)$& 1/2&1& {\bf8} &qqb&$\Sigma_b-\Lambda_b=? $&183&\\
            &$\Lambda_b(5640)$& 1/2&0& {\bf8}&qqb&$\Sigma_b^*-\Lambda_b=? $&201&\\
                             \hline
   \end{tabular}
  \]
\end{table}

\subsection{\label{Mass splittings of the flavor singlet hybrid baryons}
                   Mass splittings of the flavor singlet hybrid baryons}

If flavor octet or decuplet hybrid baryons had been identified, 
we could use them to determine the effective quark-gluon 
couplings $\kappa_{ig}$; instead we use the ansatz that 
$\Lambda_s(1405)$ and $\Lambda_s(1520)$ are flavor singlet 
hybrid baryons.
When the $\bf^21$ three quark state couples to the constituent
gluon (spin triplet), a $J=\frac{1}{2}$ and a $J=\frac{3}{2}$ hybrid state
are formed. From~(\ref{qq interaction}),~(\ref{quark-gluon decomposition}) and tables
\ref{The quark-quark interactions for color octets}
and \ref{Values for sumiSiSgFiFg} we find a mass separation of
\begin{eqnarray}
 E_{hyp}(J=3/2)- E_{hyp}(J=1/2)&=& 3  \,\kappa_{3g}.
\end{eqnarray}
For the strange and the charm system we have
\begin{eqnarray}
  \Lambda_s(1520)-\Lambda_s(1405) &=& 115~{\rm MeV} = 3 \, \kappa_{sg}\\
  \Lambda_c(2625)-\Lambda_c(2593) &=&  32~{\rm MeV} = 3 \, \kappa_{cg}.
\end{eqnarray}
As will be seen below, we could alternatively fix one difference
and predict the other.
The values for $\kappa_{sg}$ and $\kappa_{cg}$ are
\begin{eqnarray}
 \kappa_{sg} &=& 38~{\rm MeV} \\
 \kappa_{cg} &=& 11~{\rm MeV}.
\end{eqnarray}
Using these values for $\kappa_{sg}$, $\kappa_{cg}$ and
relation~(\ref{kappa relation})
we find for $i=s$ and $i=c$ respectively,
%\begin{eqnarray}
%   \begin{array}{lcccrcr}
$		\kappa_g = 61~{\rm  MeV}$ and 
$  \kappa_g = 54~{\rm  MeV} $.
%    \end{array}
% \end{eqnarray}
The similarity of these values is a check on the validity 
of the hybrid ansatz. The average  value of 
\begin{eqnarray} \label{value of kappag}
 \kappa_g &=&  (58 \pm 4)~{\rm  MeV},
\end{eqnarray}
with a relative error less then 10\% is sufficient for our effective model
to be predictive. 

Finally, to estimate the mass difference of 
$\Lambda_b(J=3/2)$ and  $\Lambda_b(J=1/2)$, we use
(\ref{kappa mass relation}), (\ref{kappa relation})
and $m_c=1.25$~GeV, $m_b=4.25$~GeV \cite{1.0} to estimate
\begin{eqnarray}
	\kappa_{bg}= \frac{m_c}{m_b} \kappa_{cg} = 3~{\rm MeV}.
\end{eqnarray}
It follows the mass difference 
\begin{eqnarray}
	\Lambda_b(J=3/2)-\Lambda_b(J=1/2) &=&  9~{\rm  MeV},
\end{eqnarray}
with nearly degenerate states because of the small value of
$\kappa_{bg}$.

\subsection{\label{Mass splittings of flavor octet and decuplet hybrid baryons}
                   Mass splittings of flavor octet and decuplet hybrid baryons}

We determine the masses and mass splittings of the flavor
octet and decuplet hybrid baryons which contain two light quarks and
one heavy quark $i$. If the hyperfine interaction $V_{hyp}$
(\ref{the interaction Hamiltonian}) would be absent, the $^{spin} flavor$
states $\bf^21$, $\bf^28$, $\bf^48$ and $\bf^210$ which form the {\bf 816}
in~(\ref{SU(18) decomposition}), would be degenerate and would have the
common mass $E_{0i}$. If $V_{hyp}$ is present, the mass of each
hybrid isospin multiplet is given by
\begin{eqnarray}   \label{asolute hybrid mass}
   E_i = E_{0i} + E_{hyp}.
\end{eqnarray}
Having found all values for 
$\kappa_{i}$, $\kappa_{g}$ and $\kappa_{ig}$,
we can calculate  $E_{hyp}=<V_{hyp}>$ 
from~(\ref{the interaction Hamiltonian}) for
$\Lambda_s(E_s=1405)$ and $\Lambda_c(E_c=2593)$
\begin{eqnarray}
E_{hyp}(\Lambda_s)=-291~{\rm MeV};\quad E_{hyp}(\Lambda_c)=-191~{\rm MeV}
\end{eqnarray}
and find with~(\ref{asolute hybrid mass}) the mean hybrid baryon masses
\begin{eqnarray}
   E_{0s}= 1696 ~{\rm MeV}; \quad E_{0c}=2784  ~{\rm MeV}.
\end{eqnarray}
We can now predict the masses of the strange and charmed flavor octet and 
decuplet hybrid baryons.
From~(\ref{qq interaction}),~(\ref{quark-gluon decomposition}) and tables
\ref{The quark-quark interactions for color octets}
and \ref{Values for sumiSiSgFiFg}
we calculate the interaction energy $E_{hyp}$ and list the
the resulting absolute masses $E_i$ (\ref{asolute hybrid mass})
of the flavor octet and decuplet in
Tab.~\ref{Masses of the flavor decuplet and flavor octet hybrids}.
We neglect a possible mixing of the hybrids with other states which
carry the same quantum numbers since a theory of mixing with nearby 
ordinary octets and decuplets must be developed first.
We also give the mass splittings $\Delta E_b$ for all the beauty
hybrids.

\begin{table}[h]
 \caption{\label{Masses of the flavor decuplet and flavor octet hybrids}
                 Predicted masses of the flavor decuplet and flavor octet
					  hybrids, energies are given in~MeV.}
	\[
   \begin{tabular}{|c|c||c|r||c|r||l|} \hline
$^{spin}flavor$& $J$ & $E_s$ & $E_{hyp}$ &$E_c$&  $E_{hyp}$ &
$\quad\quad\quad\Delta E_b$ \\ \hline\hline
$\bf^48$& $5/2$ & 1809 & 113 & 2882 &  98 & $E(5/2)-E(3/2) = 76$ \\
 (I=1)  & $3/2$ & 1689 &  -7 & 2796 &  12 & $E(3/2)-E(1/2) = 46$ \\
        & $1/2$ & 1617 & -79 & 2744 & -40 &                      \\ \hline
$\bf^48$& $5/2$ & 1792 &  96 & 2847 &  63 & $E(5/2)-E(3/2) = 123$\\
 (I=0)  & $3/2$ & 1655 & -41 & 2722 & -62 & $E(3/2)-E(1/2) = 73$ \\
        & $1/2$ & 1573 &-123 & 2647 &-137 &                      \\ \hline
$\bf^28$& $3/2$ & 1721 &  25 & 2844 &  60 & $E(3/2)-E(1/2) = 87$ \\
 (I=1)  & $1/2$ & 1634 & -62 & 2757 & -27 &                      \\ \hline
$\bf^28$& $3/2$ & 1637 & -59 & 2666 &-118 & $E(3/2)-E(1/2) = 5$  \\
 (I=0)  & $1/2$ & 1580 &-166 & 2649 &-135 &                      \\ \hline
$\bf^210$&$3/2$ & 1838 & 142 & 2884 & 100 & $E(3/2)-E(1/2) = 83$ \\
 (I=1)  & $1/2$ & 1808 & 112 & 2813 &  29 &                      \\ \hline
   \end{tabular}
  \]
\end{table}

\section{\label{Mass splittings in the mesonic sector}
                Mass splittings in the mesonic sector}

In Section \ref{Mass splittings of hybrid baryons} we
have determined the value of the effective quark-gluon coupling
$\kappa_g=60$~MeV, which explains the  
$\Lambda_s(1405)-\Lambda_s(1520)$ splitting in the baryonic sector.
As a matter of course, this value was obtained under the assumption
that $\Lambda_s(1405)$ and $\Lambda_s(1520)$ are hybrid baryons.
Although the found value also gives the correct splitting  of
$\Lambda_c(2625)-\Lambda_c(2593)$, an independent determination
of $\kappa_g$ is desirable. We want to stress that we cannot
determine $\kappa_g$ in our formalism independently of the hybrid 
assumption, since hybrids have not yet been discovered in experiment. 
However, theoretical mass predictions of hybrid mesons have been performed
in the bag model \cite{CS}. In this section we will use the 
hybrid meson mass splittings calculated in \cite{CS}, in order
to obtain a value in the mesonic sector, denoted by $\kappa_g^M$, 
and relate it to the value we have found in the baryonic sector.

In our formalism the hyperfine interaction for hybrid mesons 
is given by
\begin{eqnarray}\label{Vhypmeson}
	V_{hyp} &=&V_{q\bar q} +V_{q g}
\end{eqnarray}
where the quark-anti quark interaction is
\begin{eqnarray}\label{Vqq meson}
V_{q\bar q} &=&
	  - \kappa^M \; {\bf S}^q\cdot {\bf S}^{\bar q}\,
	  {\bf F}^q\cdot {\bf F}^{\bar q}, 
  \end{eqnarray}
with $\kappa^M$ the coupling strength between the quark and
the anti quark inside the meson, which we determine in Section
\ref{Quark-quark couplings for mesons},
and the quark-gluon interaction
\begin{eqnarray}\label{Vqg meson}
V_{qg} &=&
	  - \kappa^M_g \; {\bf S}^q\cdot {\bf S}^g \,
	  {\bf F}^q\cdot {\bf F}^g
	  - \kappa^M_g \; {\bf S}^{\bar q}\cdot {\bf S}^g\,
	  {\bf F}^{\bar q} \cdot {\bf F}^g,
\end{eqnarray}
with  $\kappa^M_g$  the coupling strength between the quark
(or anti quark) and the gluon inside the hybrid meson,
which we determine in Section
\ref{Mass splittings of hybrid mesons}. 
Compared to the effective couplings $\kappa$ found for baryons in Section 
\ref{Mass splittings of hybrid baryons}, we expect that the
values $\kappa^M$ for mesons are somewhat larger.
The reason is that more precise mass predictions can be made
using interactions which include the bag radius $r$ of the 
hadron \cite{BuFaPu,CS},
\begin{eqnarray} 
     V_{q q'} &=&
	  - \frac{\kappa}{r} \; {\bf S}^q\cdot {\bf S}^{ q'}\,
		  {\bf F}^q\cdot {\bf F}^{ q'},
\end{eqnarray}
an effect which we have neglected in this work so far due to simplicity. 
In the bag model~\cite{CS} mesons are more spatially compact than baryons,
with $r^M\approx 0.8-1.1$~fm and $r^B\approx 1.3-1.4$~fm 
and thus $r^B/r^M\approx1.3-1.6$. We have to include this scaling
if we want to relate the couplings from the mesonic sector to those of 
the baryonic sector, and take very approximately 
$\kappa^M \approx 1.6\;\kappa$.

\subsection{\label{Quark-quark couplings for mesons}
                   Quark-quark couplings for mesons}

The quark-anti quark interaction (\ref{Vqq meson}) can be written as 
\begin{eqnarray} \label{quark-anti quark interaction}
V_{q\bar q} &=&
%	- \kappa^M \; {\bf S}^q \cdot {\bf S}^{\bar q} \,
%	{\bf F}^q\cdot {\bf F}^{\bar q} \\
		  - \kappa^M \;\frac{1}{4}
                       \left[
                            ({\bf S}^{q}+{\bf S}^{\bar q})^2
                           -({\bf S}^{q})^2-({\bf S}^{\bar q})^2
                        \right]
                       \left[
                            ({\bf F}^{q}+{\bf F}^{\bar q})^2
                           -({\bf F}^{q})^2-({\bf F}^{\bar q})^2
								\right],
\end{eqnarray}
%where $\kappa^M$ is the coupling strength between the quark and
%the anti quark inside the meson, which we will determine in this
%section.
and the splitting of the $S=J=1$ and the $S=J=0$ meson 
multiplet is thus given by
\begin{eqnarray} 
	E(J=1)-E(J=0)=\frac{4}{3}\kappa^M_i.
\end{eqnarray}
The effective couplings $\kappa^M,\kappa^M_s,\kappa^M_c,\kappa^M_b$ 
for mesons can now be determined,
see table \ref{The mesons used to determine kappai}.
\begin{table}[h]
   \caption{\label{The mesons used to determine kappai}
                The mesons used to fix the $\kappa^M_i$}
	\[
   \begin{tabular}{|l|c|c|c|c|c|c|c|} \hline
  $\kappa_i^M$&  meson&J&content&
  $\Delta M~[{\rm MeV}]$&$\kappa^M_i~[{\rm MeV}]$
  &$\kappa_i=\kappa^M_i/1.6~[{\rm MeV}]$
  \\ \hline\hline
  $\kappa^M$ &${\rho(776)\atop\pi(138(280)) }$&
  ${1\atop0}$&qq&638(496)&479(372)&299(233)   \\
  \hline
  $\kappa^M_s$ &${K^*(894)\atop K(496) }$& ${1\atop0}$&qs&398&299&187   \\
  \hline
  $\kappa^M_c$ &${D^*(2007)\atop D(1865) }$& ${1\atop0}$&qc&142&107&67   \\
  \hline
  $\kappa^M_b$ &${B^*(5325)\atop B(5279) }$& ${1\atop0}$&qb&46&35&22   \\
  \hline
   \end{tabular}
  \]
\end{table}
Note that we obtain a more reasonable value for $\kappa^M=372 $~MeV if we take
the predicted mass of $\pi_{bag}(280)$ in the bag model \cite{CS}, 
instead of the physical mass $\pi(138)$. In the limit of chirality 
conservation the $\pi$ is the 'would be Goldstone boson' and
thus has a light mass. The bag model cannot account for this effect,
thus we will use $\kappa=372 $~MeV. Comparing the $\kappa_i$ with the 
values obtained from baryon splittings, see 
Table \ref{The particles used to determine kappai} in 
Section \ref{Quark-quark couplings for baryons}, we find quite good agreement, 
since we included the size effect of the radii.

\subsection{\label{Mass splittings of hybrid mesons}
                   Mass splittings of hybrid mesons}

We consider hybrid mesons consisting of two quarks $q$ and $\bar q$
which are light $q=u,d$. We will also allow that one of the
quarks can be strange $q=s$. The two quarks form a color 
octet state with either $S=0$ or 
$S=1$ which couples to a TE gluon with $J^{PC}=1^{+-}$.
The $q\bar q$ state with $S=0$ makes then a $J^{PC}=1^{--}$ hybrid 
meson, and the $S=1$ $q\bar q$ state makes three 
hybrid mesons  with $J^{PC}=0^{-+},1^{-+},2^{-+}$. 
Due to the hyperfine interaction~(\ref{Vhypmeson}) the degeneracy 
of these hybrid meson multiplets is lifted. 
We give our calculated values in  Table \ref{mesons}.
\begin{table}[h]
  \caption{\label{mesons}
                Summary table of hybrid meson interaction energies}
	\[
   \begin{tabular}{|c|c|c|c|c|} \hline
 $J^{PC}$ &     type & content & $<V_{q\bar q}>$ & $<V_{qg}>$ \\ \hline\hline
 $1^{--}$ & $\rho/\omega$& qq  & $1/8 \kappa^M$    & 0 \\ \hline
 $1^{--}$ & $ K^* $      & qs  & $1/8 \kappa^M_s$  & 0 \\ \hline \hline
 $0^{-+}$ & $\rho/\omega$& qq  & $-1/24 \kappa^M$  & $-3\kappa^M_g $\\ \hline
 $0^{-+}$ & $ K^* $      & qs  & $-1/24\kappa^M_s$ &
		$-3/2(\kappa^M_g+\kappa^M_{gs}) $ \\ \hline 
 $1^{-+}$ & $\rho/\omega$& qq  & $-1/24 \kappa^M$  & $-3/2\kappa^M_g $\\ \hline
 $1^{-+}$ & $ K^* $      & qs  & $-1/24\kappa^M_s$ &
		$-3/4(\kappa^M_g+\kappa^M_{gs}) $ \\ \hline 
 $2^{-+}$ & $\rho/\omega$& qq  & $-1/24 \kappa^M$  & $3/2\kappa^M_g $\\ \hline
 $2^{-+}$ & $ K^* $      & qs  & $-1/24\kappa^M_s$ &
		$3/4(\kappa^M_g+\kappa^M_{gs}) $ \\ \hline
     \end{tabular}
  \]
\end{table}

The absolute masses of the non-strange hybrids $\rho/\omega$
and the strange hybrids $ K^*$ have been calculated in \cite{CS} 
for three different values of the ratio of the gluon $TE$ and $TM$ 
self energies, $C_{TE}/C_{TM}=\frac{1}{2},1,2$. We obtain the same 
ordering of states for each of these ratios.
We use the splittings of $\rho/\omega$ to determine $\kappa^M_g$
and find $100~{\rm MeV} \lsim \kappa^M_g \lsim 140~{\rm MeV}$.
Fitting the splittings of the strange hybrids  $ K^*$ 
we obtain $60~{\rm MeV}\lsim\kappa^M_{gs}\lsim130~{\rm MeV}$.
These values of the coupling strengths are consistent with the results 
from ordinary meson splittings, since they have to obey $(\ref{kappa relation})$
\begin{eqnarray}
  \frac{\kappa^M}{\kappa_g^M} &=& \frac{\kappa_s^M}{ \kappa_{gs}^M},
	\end{eqnarray}
which is quite well fulfilled for our values
$\kappa^M=370$~MeV, $\kappa^M_g=(120\pm20)$~MeV and
$\kappa^M_s=300$~MeV, $\kappa^M_{gs}=(95\pm35)$~MeV.
%which we have obtained from the mesonic sector.
Deducing from 
$100~{\rm MeV}\lsim\kappa^M_{g}\lsim140~{\rm MeV}$
a value for the baryonic sector, we include the scaling 
$\kappa^M_g \approx 1.6\;\kappa_g$,
as discussed in the beginning of this section, and find
$60~{\rm MeV}\lsim\kappa_{g}\lsim90~{\rm MeV}$.
This value is compatible with our result $\kappa_{g}=60~{\rm MeV}$ 
obtained in (\ref{value of kappag}), and adds credibility to the hypothesis 
that $\Lambda(1405)$ and $\Lambda(1520)$ are hybrid baryons.

\section{\label{The parity of Lambda(1405)}
                The parity of $\Lambda(1405)$}

As noted in the Introduction, the parity of $\Lambda(1405)$ has not
been directly measured experimentally \cite{Dal,Hem,Thom}.
This is because the
$\Lambda(1405)$ mass is below threshold in $KN$ scattering, and
it must be studied in $\Sigma \pi$ scattering, specifically in $K p
\rightarrow \Sigma \pi \pi \pi$ via $\Lambda^* \rightarrow \Sigma \pi$
followed by $\Sigma \rightarrow p \pi$, where $\Lambda(1405)$ is
observed as a resonance in the $\Lambda^*$ channel.  Unfortunately,
the dynamics of the $\Lambda^*$ production in this process are such
that it is produced with a small degree of polarization, so the
interference between $s$- and $p$-wave final states which provides
the sensitivity to the relative $\Lambda^*-\Sigma$ parity is small and
the parity is therefore very difficult to measure \cite{Hem}.  In
principle this might be rectified by an experiment such as $K p
\rightarrow \Sigma \pi \pi \gamma$, where there is at least the
opportunity for more favorable production dynamics. Additionally, if
the hybrid interpretation of the $\Lambda(1405)$ is correct, it might
be {\it comparatively} copiously pair produced in $J/\Psi$ decay.  The
conclusion of \cite{Dal} that $\Lambda(1405)$ has negative parity is
based on an indirect argument about the below-threshold line shape.
Argumentation based on dynamical expectations to distinguish between
$s$- and $p$-wave amplitudes is particularly dangerous because no
known model fits the energy dependence of the data.  As noted
by Dalitz \cite{Dal}, the argument for negative parity is very weak and
it is only adopted because there seemed no theoretical motivation to
consider the parity to be positive.

The {\it lightest} hybrid flavor singlet baryon is likely to
have even parity because the bag model predicts \cite{Closebook}
the lightest $J^P=1^{(-)} (TM)$ gluon mode is about $300$~MeV
heavier than the lightest $J^P=1^{(+)} (TE)$ gluon mode.  This
is also corroborated by lattice calculations \cite{Juge}.
Therefore if the $\Lambda(1405)$ and $\Lambda(1520)$ are the
lightest flavor singlet hybrid baryons they should have even parity.
Their negative parity partners should have masses of about 
$1.7$~GeV and $1.8$~GeV.  The
experimental detection of such resonances would be difficult,
due to mixing with ordinary quark model states which are
abundant in that mass range. The second possibility is that the
$\Lambda(1405)$ and $\Lambda(1520)$ have odd parity and are {\it
not} the lightest flavor singlet hybrid baryons.  This would
imply the existence of a lower mass pair of even parity states
with mass about $1.1$~GeV and $1.2$~GeV! It is doubtful to us that such
low energy states could be discovered. They would be below
threshold in any baryon number $B=+1$, strangeness $S=-1$
meson-baryon state so their effect could only be observed as a
resonance in $N \pi$, or possibly $\Lambda \gamma$ if the
resonance mass is heavy enough. However the effect of a light strange
even-parity hybrid resonance on a $\pi N$ invariant mass plot or
partial wave analysis would be extraordinarily small because it
would be coupled only via weak interaction and only effect
$p$-wave states. The most promising place to look might be in
$J/\Psi$ decays, where hybrid baryon pair production might be
favored by OZI considerations; if kinematically allowed,
$\Lambda \gamma$ would be the most promising final state.

It is astonishing to contemplate that one or more new
strong-interaction-stable baryon states might exist, but we can see no
compelling argument against it.  Fortunately, this possibility has other
implications which can be tested, as will be discussed in the next
section.  Lattice QCD calculations of masses of the positive parity
flavor singlet baryon spectrum would be most illuminating in this
context, and should be feasible.  If they are consistent with the
masses under discussion here ($1.1$~GeV - $1.4$~GeV), every effort should be
made to pursue the experimental searches suggested above. In the
absence of additional information favoring this hypothesis, we favor
the more conservative hypothesis that the $\Lambda(1405)$ and 
$\Lambda(1520)$ have positive parity. 

\section{\label{A low lying dihyperon ?}
                A low lying dihyperon?}

We have explored the ansatz that a $uds$ in a color octet,
flavor singlet state binds with a constituent gluon to produce
the $\Lambda(1405)$. We have analyzed mass splittings between
members of the various multiplets, but have made no absolute mass
predictions.  Lattice QCD would be the only trustworthy way to
do this because \emph{absolute} mass predictions of
phenomenological models such as Skyrme, MIT bag or potential
models are notoriously unreliable.  To the best of our
knowledge, no lattice calculation of hybrid baryon masses has
yet been performed.
However, glueball masses have been calculated on the
lattice \cite{gluelatt}.  The lightest (pure) glueball is
predicted to have a mass in the range $1.4$~GeV - $1.7$~GeV and there
are good glueball candidates in this range. If the hybrid baryon
ansatz is accepted for the $\Lambda(1405)$ and $\Lambda(1520)$,
then the approximate coincidence of the $udsg$ and $gg$ masses
implies that a color octet state of three light quarks in a
flavor singlet is approximately equivalent to a gluon as a
dynamical system.  This is not an outlandish idea as the dynamics
of a hadronic bound state depends primarily on the color, mass
and spin of the constituents.  Thus a spatially-compact $uds$
system in a flavor singlet state naturally resembles a gluon
just having spin 1/2 rather than spin 1. Given the uncertainty
discussed above as to whether or not the $\Lambda(1405)$ and
$\Lambda(1520)$ are the lightest states, the uncertainty
in``constituent" mass of this flavor-singlet, color-octet $uds$
system is 300-500~MeV.

Pursuing the ansatz that a flavor singlet $uds_{\bf 8}$ system behaves
much like a gluon, a combination of two $uds_{\bf 8}$ should be a
glueball-like state with mass below $\sim 2$~GeV. This would be
the H-dibaryon, which is an even parity six-quark state with
spin $J = 0$ and isospin $I = 0$, baryon number $B = 2$ and
strangeness $S=-2$. The H was predicted in 1977 by Jaffe
\cite{Jaffe} in a MIT bag model calculation, which estimated the mass
to be about $2150$~MeV. Since then, many other H mass calculations
have been performed, using Skyrme, quark cluster models, lattice 
calculations and instanton-based interactions \cite{kochelev:H}
(for an overview see \cite{Hdib}).  The resulting
mass estimates range from less than $1.5$~GeV to more
than $2.23$~GeV ($\Lambda \Lambda$-threshold) which would make
the H decay strongly. The wide range of the mass predictions for
the H reflect the theoretical uncertainties in its existence and
structure. If the $\Lambda(1405)$ is a hybrid baryon, we suggest
that the mass of the H-dibaryon could lie in the $1.5$-$2$~GeV mass
region\footnote{See \cite{f:stableH} for a more refined mass
estimation.}. If the H is lighter than two nucleons, 
an important concern would be the
stability of nuclei against conversion over cosmological times,
of pairs of nucleons to the H. Remarkably, this and other
phenomenological difficulties with a stable or long-lived H are
not fatal, if the H is sufficiently compact \cite{f:stableH,
fz:nucStab, fz:binding}. The discovery of a light H, especially
if its mass were such that it was the ground state for $B=2$,
would be a stunning discovery in its own right. As a byproduct,
it would lend strong support to the proposal that the
$\Lambda(1405)$ is a hybrid. On the other hand a stable H is
not a necessary consequence of the hypothesis of a strong
attraction in the $uds_{\bf 8}$ system, so excluding it would
not be sufficient to exclude the suggestion that
$\Lambda(1405)$, etc, are hybrids. Lattice QCD calculations will
be the most effective tool for that.

\section{\label{Summary and Conclusion}
                Summary and conclusions}

We have explored the hypothesis that a very strong attraction in
the $uds_{\bf 8}$ state may mean that the four isosinglet
baryons $\Lambda_s(1405)$, $\Lambda_s(1520)$, $\Lambda_c(2593)$
and $\Lambda_c(2676)$ are hybrids.  The observed mass splittings
are consistent with the hybrid baryon hypothesis, resolving a
persistent problem of the conventional identification as an
orbital excitation of a 3-quark state.  It is non-trivial that
the ordering of states is $m_{J=3/2}>m_{J=1/2}$, as observed
experimentally, because in the conventional $L=1$ picture the $J=3/2$
state is necessarily the lightest due to spin-orbit interactions. 
Assuming these states are flavor singlet hybrid baryons fixes the parameters of the
quark-gluon hyperfine interaction. We have shown that the obtained
quark-gluon coupling is consistent with calculations from the 
mesonic sector. In addition, this allows the mass
splittings of the flavor octet and decuplet hybrid baryons to be
predicted.  A theory of mixing with nearby ordinary octets and
decuplets must be developed before these predictions can be
tested.

The best experimental test of our ansatz that the
$\Lambda_s(1405)$, $\Lambda_s(1520)$, $\Lambda_c(2593)$ and
$\Lambda_c(2676)$ are hybrids, is to determine whether they are
parity doubled, with the odd parity partner about $300$~MeV
heavier than the even parity state.  This scenario predicts
either that the $\Lambda_s(1405)$, $\Lambda_s(1520)$,
$\Lambda_c(2593)$ and $\Lambda_c(2676)$ are even parity, or that
there are as-yet-undiscovered even parity flavor singlet,
strangeness $-1$ states at about $1.1$~GeV and $1.2$~GeV.  The hybrid
ansatz suggests, but does not predict, that the H-dibaryon
mass may be as low as $1.5$~GeV.  Further work is needed to see if
these possibilities can be excluded on observational grounds. In
parallel, lattice QCD calculation of the masses of $udsg$ states
and of the H-dibaryon mass will indicate whether this is a
correct interpretation.

\section{Acknowledgement}

This work was supported in part by NSF-PHY-99-96173 and NSF-PHY-0101738.

\appendix

\section{\label{Casimir operators}
                Casimir operators}

Values for the $SU(2)$ and $SU(3)$ Casimir operators
\begin{eqnarray}
	C_2={\bf S} \cdot {\bf S} = \frac{1}{4}\sum_{m=1}^3 \sigma_m\sigma_m, 
		\quad \quad
	C_3= {\bf F} \cdot {\bf F} = \frac{1}{4}\sum_{a=1}^8 \lambda_a\lambda_a, 
\end{eqnarray}
are given by:
	\[
  SU(2):
           \begin{array}{|c||c|c|c|c|c|} \hline
            dim &{\bf1}&{\bf2}&{\bf3}&{\bf4 }&{\bf5} \\ \hline
            C_2 &  0   &  3/4 &  2   &  15/4 &     6 \\ \hline
           \end{array}
          \;\;\;\;\;
   SU(3):
           \begin{array}{|c||c|c|c|c|c|} \hline
             dim &{\bf1}&{\bf3}&{\bf6}&{\bf8}& {\bf10} \\ \hline
             C_3 &  0   &  4/3 & 10/3 &   3  &  6      \\ \hline
           \end{array}
   \]

\section{\label{Mixed symmetry functions}
                   Mixed symmetry functions}
					 
\begin{table}[h]
	\caption{\label{The functions} The definitions of mixed symmetric and  mixed 
		antisymmetric wave functions $\varphi_{MS}$ and $\varphi_{MA}$, 
		respectively. The table is taken from \cite{Closebook}.}
	\begin{tabular}{|c|c|c|} \hline
  $label$  &  $\varphi_{MS}   $   &  $\varphi_{MA}  $ \\ \hline \hline
  $P$  &  $   \frac{1}{\sqrt{6}}
             \left[ (ud+du)u -2uud
             \right]$
        &  $\frac{1}{\sqrt{2}}(ud-du)u  $ \\ \hline
 $N$  &  $  - \frac{1}{\sqrt{6}}
             \left[ (ud+du)d -2ddu
             \right]$
 &  $\frac{1}{\sqrt{2}}(ud-du)d  $ \\ \hline
 $\Sigma^+$ &  $   \frac{1}{\sqrt{6}}
             \left[ (us+su)u -2uus
             \right]$
&  $\frac{1}{\sqrt{2}}(us-su)u  $ \\ \hline
 $\Sigma^0$ &$  \frac{1}{\sqrt{6}}
           \left[ s \left( \frac{du+ud}{\sqrt{2}} \right)
                  + \left( \frac{dsu+usd}{\sqrt{2}} \right)
          \right. $
 &  $\frac{1}{\sqrt{2}}
          \left[   \left( \frac{dsu+usd}{\sqrt{2}} \right)
                -s \left( \frac{ud+du}{\sqrt{2}}  \right)
            \right]$ \\
 & $ \left.
          -2 \left( \frac{du+ud}{\sqrt{2}} \right)s
  \right]$ &  \\ \hline
$\Sigma^-$ &  $  \frac{1}{\sqrt{6}}
             \left[ (ds+sd)d -2dds
             \right]$
&  $\frac{1}{\sqrt{2}}(ds-sd)d  $ \\ \hline
$\Lambda^0$ &  $\frac{1}{\sqrt{2}}
          \left[   \left( \frac{dsu-usd}{\sqrt{2}} \right)
                +         \frac{s(du-ud)}{\sqrt{2}}
         \right]$
 &$   \frac{1}{\sqrt{6}}
         \left[     \frac{s(du-ud)}{\sqrt{2}}
                +   \frac{usd-dsu}{\sqrt{2}}
         \right.   $ \\
&&$ \left. -\frac{2(du-ud)s}{\sqrt{2}} \right]$ \\ \hline
$\Xi^-$ &  $ - \frac{1}{\sqrt{6}}
             \left[ (ds+sd)s -2ssd
             \right]$
 &  $\frac{1}{\sqrt{2}}(ds-sd)s  $ \\ \hline
$\Xi^0$ &   $ - \frac{1}{\sqrt{6}}
             \left[ (us+su)s -2ssu
             \right]$
 &  $\frac{1}{\sqrt{2}}(us-su)s  $ \\ \hline
 \end{tabular}
\end{table}

\appendix

\end{document}